
\font\cmss=cmss10 \font\cmsss=cmss10 at 7pt
\def\dddots{\mathinner{\mkern2mu\raise1pt\vbox{\kern7pt\hbox{.}}\mkern2mu
   \raise4pt\hbox{.}\mkern2mu\raise7pt\hbox{.}\mkern1mu}}

\def\IR{\relax{\rm I\kern-.18em R}}
\def\IN{\relax{\rm I\kern-.18em N}}
\def\IC{\relax\thinspace\hbox{$\inbar\kern-.3em{\rm C}$}}
\def\IZ{\relax\ifmmode\mathchoice
   {\hbox{\cmss Z\kern-.4em Z}}{\hbox{\cmss Z\kern-.4em Z}}
   {\lower.9pt\hbox{\cmsss Z\kern-.4em Z}}
   {\lower1.2pt\hbox{\cmsss Z\kern-.4em Z}}\else{\cmss Z\kern-.4em Z}\fi}

\overfullrule=0pt
\hsize=6.0in
\hoffset=0.0in
\voffset=0.0in
\vsize=8.9in
\FRONTPAGE
\line{\hfill BROWN-HET-869}
\line{\hfill August 1992}
\vskip1.5truein
\titlestyle{{INTERACTING THEORY OF
COLLECTIVE\break
AND TOPOLOGICAL FIELDS IN 2 DIMENSIONS}}
\bigskip
\author{Jean AVAN\foot{Address from Sept. 1$^{st}$, 1992:  LPTHE
Paris 6 (CNRS-UA 280), Box 126, 4 place Jussieu, F-75252 Paris Cedex
05} and Antal JEVICKI}
\centerline{{\it Department of Physics}}
\centerline{{\it Brown University, Providence, RI 02912, USA}}
\bigskip
\abstract

\noindent

{\bf \chapter{\bf Introduction}}

Recent investigations of matrix models and two dimensional strings
(for a review see [1]) have lead to some deeper understanding of the
theory and its fundamental symmetries.  The basic space--time picture
was given by a bosonic (collective) field representing the main
degree of freedom (the massless tachyon [2-4]).  Generally a matrix
model with Hamiltonian $\hat{h} = Tr (P^2 + U(M)) $ is equivalent to a
$N$--body Fermi
gas with Hamiltonian $H_N = \sum_{i=1}^N \bigl( - {\partial^2\over
\partial\lambda_i^2}
+ U (\lambda_i )\bigr)$ which is described by a single nonrelativistic
Fermi field $\Psi (x)$ [5,6].

The bosonic (collective field) is
$$\phi (x) = \Psi^{\dagger} (x) \Psi (x)\, ; \, \Pi(x) \,\, {\rm is
\,\, its \,\,conjugate\,\,momentum}\eqno\eq$$
with a nonlinear hamiltonian
$$H_{{\rm coll}} = \int dx \left\{ {1\over 12\pi} \right( \alpha_+
(x,t)^3 - \alpha_- (x,t)^3 \left) + {1\over2\pi} U (x) \left( \alpha_+
- \alpha_- \right)\right\}\eqno\eq$$
and $\alpha_{\pm} = \Pi_{,x} \pm \phi$ are the left (right) moving
fields.  The presence of a nontrivial cubic interaction results in
scattering processes for the bosonic excitations [4].

The symmetry generators [7,8] are given by polynomials
$$H_m^n = Tr \left( M^m P^m \right) = \int dx {x^n\over m+1} \left(
\alpha_+^{m+1} - \alpha_-^{m+1} \right)\eqno\eq$$
which build a $w_{\infty}$ algebra [9], obeying the Poisson brackets:
$$\left\{ H_{m_1}^{n_1} , H_{m_2}^{n_2} \right\} = \left( n_1 m_2 -
n_2 m_1 \right)
H_{m_1 + m_2}^{n_1 + n_2}\eqno\eq$$
One then constructs the spectrum generating operators
$\hat{W}_{JM} = Tr \left( (P+M)^{J+M}\nextline
 (P-M)^{J-M} \right)$ which produce an infinite sequence of imaginary
energy discrete
states [7,10].

These and possibly more general features also arose in
the first quantized conformal field theoretic description.  Here the
vertex operators of discrete states [11] acting on the ground ring of
observables were
shown to close under operator products resulting again in a
$w_{\infty}$ algebra [12].  A field theoretic formulation was
suggested where all the vertex operators are represented by separate
fields and a cubic coupling of the discrete fields was given [12].
Extensions in various directions of the collective field theory were
also
considered [13,14,15].

The interplay of the tachyonic and discrete degrees of
freedom is still however a considerable mystery.  Most importantly the
complete theory describing both tachyon and discrete fields is yet not
given.

In the present paper we address these problems.  Our discussion is
done at the quantum level.  Of basic relevance for
understanding the combined theory will be the notion of target
space--time (as opposed to world--sheet) central charges.  These
charges which we will associate with the space--time fields
(collective and discrete) appear once one looks at a chiral
decomposition into left and right--moving fields.  In the combined
algebra the central charges cancel between the two chiralities.
Along these lines we analyze the most general
representations of the basic quantum $W_{1+\infty}$ algebra.

We derive the following physical picture:  in addition to the tachyon
$\{\alpha_{\pm}\}$ one has an infinite set of discrete fields
$\{w_n^{\pm}
\}$.  We show that the two sets of fields interact with a
{\it unique} coupling, the complete theory being given by the
hamiltonian:
$$ H \equiv \sum_{(+,-)}  \int dx   \left[ {1\over 12\pi}
\alpha(x,t)^3 + \alpha(x,t) w_1
(x,t) + w_2 (x,t) \right]^{(+-)}\eqno\eq$$
and the algebraic structures:
$$\eqalign{
& \left[ \alpha_{\pm} , \alpha_{\pm} \right] = \pm 2\pi \partial_x
\delta (x-y)\cr
& \left[ w_m (x) , w_n (y) \right] = \left\{ (n-1) w_{m+n} (x) + (m-1)
w_{m+n} (y)\right\} \partial_z \delta (z-z')
}\eqno\eq$$

After quantization the scalar (collective) field $\alpha_{\pm}$ is
determined so as to carry {\bf all} the central charge $c=1$ while the
discrete fields $w_n$ are constrained to have $c=0$ and carry
therefore essentially no degree of freedom.  We then have a
theory of a collective boson interacting with a set of discrete
{\bf topological} fields.  The latter can be represented either as a
three--dimensional system derived from Toda theory [16] which at
quantum level should possibly be
amended with cancelling ghost fields; or as functionals on a coadjoint
orbit of the algebra of pseudo--differential operators of KP type,
written as total derivatives.  These two schemes are in fact two
parametrizations of the large $N$ limit of the KdV hierarchy.  This
approach is to be contrasted with other treatments where all fields
are considered on the same footing.

The additional discrete fields have some clear physical relevance
which we discuss in the last section of the paper.  They enter
through the boundary
conditions and in principle can serve as nontrivial classical
backgrounds.

The content of the paper goes as follows.  In section 2 we consider
the algebra of fermion bilinear observables and its representations.
It is shown how the generators (of the $W_{1+\infty}$ algebra) are
decomposed into a single boson (collective field) contribution and
additional discrete contributions.  The Hamiltonian is then written in
terms of these variables.  In section 3 we discuss field theoretical
representations.  The topological representation is emphasized.  In
section 4 we exemplify the physical relevance of the scheme.
Appendixes are reserved for more technical considerations.

\endpage

{\bf\chapter{\bf The Bilinear Fermion Algebra}}

We start with a set of non--relativistic fermions in 1 space
dimension with a simple
Hamiltonian of the form:
$$\eqalign{ {\cal H}  & = \int dx \left( {1\over 2} \partial_x
\Psi^{\dagger} \partial_x \Psi + U (x) \Psi^{\dagger} \Psi \right)\cr
\Psi^{(\dagger )} & = {1\over \sqrt{L}} \, \sum_{n=-\infty}^{+\infty}
\, e^{2\pi in/L} \, C_n^{\dagger }\cr
 & \left[ C_n , C_m^{\dagger} \right]_+ = \delta_{n,m} }\eqno\eq$$
We then consider a general set of bilinear observables:
$$O^{K} (x) = \Psi^{\dagger} (x) \, \partial_x^K \, \Psi
(x) \eqno\eq$$
rewritten in dual Fourier space as:
$$O_n^K = \sum_{m = -\infty}^{+\infty} \, m^K \, C_m^{\dagger}\,
C_{m+n}\eqno\eq$$
They are separated into two subsets of operators, as:
$$\eqalign{ O_n^{K} & = O_{+}^{K }_n \, + O_{-}^{K}_n \cr
O_{\pm}^{K} _n & = \sum_{m_<^> -{n\over 2}} \, m^K \,
C_{m+n}^{\dagger} \, C_n }\eqno\eq$$

The Fermi surface of these non--relativistic free fermions is given by
the two Fermi momenta $\pm p_F$.  Near these momenta the excitation
spectrum becomes linear.  The above $\pm$ components create
excitations near the $\pm p_F$ values respectively.

One can then show that these two sets decouple when acting on the
physical Hilbert space:
$$\left[ O_+^{K_1 } _{(m_1)} , O_-^{K_2} _{(m_2)} \right] \vert_{\rm
Hilbert} = 0\eqno\eq$$

This separation originally due to Tomonaga [17] can be
reformulated in terms of two respectively left
-- and right -- moving Dirac fermions:
$$\Psi(x) \equiv e^{i\pi/4} \, \psi_+ (x) + e^{-i\pi/4} \, \psi_-
(x)\eqno\eq$$

We concentrate from now on on the two decoupled branches
$\pm$ separately, and we shall carry in what follows the notation for
only one Dirac fermion chirality.  In the final result, the $+$ and $-
$ contributions will naturally be added.

\section{The Algebra of Relativistic Fermions}

The Dirac fermion fields $\psi^\dagger (z)$ and $\psi (z)$ are
expanded in terms of creation and annihilation operators as:
$$
\psi^\dagger (z) = \sum_{r \epsilon\IZ + 1/2} \psi_{-r}^\dagger
z^{r-1/2}
\quad ; \quad \psi (z) = \sum_{r\epsilon \IZ + 1/2} \psi_r z^{r-
1/2}\quad ; \quad \left[ \psi_r^{\dagger} , \psi_s\right]_+ =
\delta_{r,s}\eqno\eq$$
Normal ordering is defined by:
$$\colon \psi_r^\dagger
\psi_s\colon \, \equiv  \psi_r^\dagger \psi_s (s>0) \,\,{\rm or}\,\,
- \psi_s \psi_r^\dagger (s<0)   \, . \eqno\eq$$
or equivalently $\colon\>\psi_r^\dagger \psi_s\>\colon \> \equiv
\psi_r^\dagger \psi_s + < \psi_r^\dagger\, \psi_s\> >,
<\> >$ denoting the vacuum
expectation value in the Fermi vacuum defined by $\psi_r
\vert\, 0\, >\> = 0
\,(r>0)\, ;\, \psi_r^\dagger\> \vert\, 0 \, >\> = 0\, (r<0)$.

We define a complete set of independent generators for the fermion
bilinears, defined to be:
$$
B_n^K = \sum_{q\epsilon\IZ}  \quad \colon
\psi_{q-n-1/2}^\dagger \> q^K
\> \psi_{q- 1/2}\colon \eqno\eq
$$
Their commutators are:
$$\eqalign{
\left[ B_{n_1}^{K_1}\, ,\, B_{n_2}^{K_2} \right ] = &\sum_q  \psi_{q-
n_1 - n_2 -1/2}^\dagger\, \left\{ (q-n_2)^{K_1} q^{K_2} -
(q-n_1)^{K_2} q^{K_1}\right\} \psi_{q-1/2}\cr
&- \delta_{n_1 + n_2} \sum_{\ell = 0}^{n_1 - 1} \ell^{K_1} (\ell -
n_1)^{K_2}\, .} \eqno\eq
$$
defining a $W_{1+\infty}$ algebra [9,18].  The $c$--number term
represents an anomaly, and plays a capital role in the subsequent
developments.
The description in [18] is given in term of an overcomplete set of
generators:
$$
W_n^{(K)} (j) = \ointop {dz\over {2\pi i}} z^{n + K-1} \quad :\,
\partial^j  \psi^\dagger (z) \partial^{n-1-j} \psi (z):\eqno\eq
$$
Here $j$ is an essentially redundant variable.  Indeed,
one can rewrite (2.11) as:
$$\eqalign{
W_n^K (j) &= \sum_q  {(q+n+1)!\over (q+n-j)!}
{q!\over (q-K+j)!} \psi_{q+n}^\dagger \psi_q\cr
&= \sum_q  q^{K-1} \psi_{q+n}^\dagger \psi_q + \sum_{j=1}^{K-1}
c_{j, K}
\left ( \sum_q  q^{K-j-1}\right ) \psi_{q+n}^\dagger \psi_q}\eqno\eq
$$

The standard bosonization of this algebra is given in term of a
single field
$$
\partial \varphi (z) = \colon \psi^\dagger (z) \psi
(z)\colon \equiv \alpha (z)\eqno\eq
$$
or conversely $\psi^\dagger (z) =\> \colon e^{\varphi (z)}$ and $
 \psi (z) =\> \colon e^{- \varphi(z) }\colon\,$. This gives a non--
linear realization since the generating currents $W(z)$ have the
generic form:
$$
W^{(n)} (z) = \quad \colon \> \alpha^n (z) \colon\> +\> {\rm lower\>
orders} \eqno\eq
$$
hence $ \Delta\alpha(z)\equiv \left\{ W^{(n)} (z),\> \alpha (z)
\right\}_{{\rm Lie}} $ is a
non-linear transformation of the field $\alpha (z)$.  The classical
limit of this particular non--linear representation is the Poisson
algebra of classical observables for the collective field theory [7].

We now want to show that the above algebra (2.10) has a more general
representation in terms of an extended set of bosonic fields.
Let us consider the
bilinear commutators in more detail.

The first commutators read:
$$\eqalign{
[B_m^0 , B_n^0 ] &= m\> \delta_{m+n} \quad ; \quad [B_m^1,
B_n^0 ] = - n \>B_{m+n}^0 + {m(m-1)\over {2}} \delta_{n+m}\cr
[B_m^2, B_n^0 ] &= n^2\> B_{n+m}^0 - 2n\> B_{m+n}^1 + {m(m-1) (2m-
1)\over {6}} \delta_{m+n}\cr
[B_m^3, B_n^0 ] &= - n^3 \> B_{n+m}^0 + 3n^2 \> B_{n+m}^1 - 3n
\>B_{n+m}^2 + \delta_{m+n}\> (\dots )\cr
[B_m^1, B_n^1 ] &= (m-n) \>B_{m+n}^1 - {m^3 - m\over {6}}
\delta_{m+n}\cr
[B_m^2, B_n^1 ] &= (m-2n) \>B_{m+n}^3 + n^2\> B_{m+n}^1 + \delta_{m+n}
\>(\dots )\cr
[B_m^3 , B_n^1 ] &= (m -3n) \> B_{m+n}^4 + 3n^2\>B^3_{m+n} \>
B_{m+n}^1 + \delta_{m+n} (\dots )\cr
[ B_m^2 ,  B_n^2 ] &= 2 (m-n) \> B_{m+n}^3 + (n^2 - m^2 ) \>B_{n+m}^2
+ \delta_{m+n} (\dots )}\eqno\eq
$$
Once the central terms are diagonalized, in particular by redefining
${\tilde B}_m^1$ as $B_m^1 - {m-1\over 2} B_m^0$, (2.15) becomes the
$W_{1 + \infty}$ algebra with central charge $c = + 1$, the central term
in $\left [ {\tilde B}_m^1 , {\tilde B}_n^1 \right ]$ becoming $+ {m^3
- m\over {12}} \delta_{n+m}$.

First of all we emphasize that the commutator for $B_m^0$
characterizes these operators as arising from a $U(1)$ current.
We can thus identify confidently the components
$B_m^0 \equiv \alpha_m$ with the operator $\alpha_m \equiv \oint z^{m-
1} \alpha(z) dz$, where $\alpha(z)$ is the basic collective degree of
freedom (tachyon), realizing indeed a $U(1)$
current algebra.

Consider next the commutator $\bigl[ B_m^1 , B_n^0\bigr]$.  The idea
is (following [17]) to solve for $B_m^1$ so that the commutator
is obeyed.  The right
hand side is given in terms of $B_n^0 = \alpha_n$ and one write
for $B_m^1$ a most general form reading:
$$B_m^1 = {1\over 2} \sum_{\ell} \colon \alpha_{m-\ell} \,
\alpha_{\ell} \colon + {m-1\over 2} \, \alpha_m +
\bar{W}_m^1\eqno\eq$$
The first two terms correspond to standard bosonization $B^1 (\alpha
)$ while $\bar{W}_m^1$ is now an (abstract) additional spin--2 field.
It commutes with $\alpha$, and its self--commutator yields a
Virasoro algebra with central charge $c=0$:
$$\left[ \bar{W}_m^1 , \bar{W}_n^1\right] = (n-m)
\bar{W}_{n+m}^1\eqno\eq$$
We then extend the above construction to higher spin fields with the
conjecture that the full $W_{1+\infty}$ quantum fermionic algebra
can be
represented in terms of a  $1$--field set of currents (namely the
$1$--boson
representation) plus higher--spin fields.  We now make the stronger
assumption that the high spin fields can be added one by one, namely
there exists a ``nested" construction of $W_{1+\infty}$ obtained by
adding at each level \underbar{one} ``residual" field $\bar{W}_s$ of
spin $s$ and computing the -- assumed consistent -- corrections to the
expressions of the higher--spin generators of $W_{1+\infty}$
previously computed up to $\bar{W}_{s-1}$.

The particular results previously obtained for $\bar{W}^1$ and
$\alpha$ will now be extended to higher spin fields.

First of all, $W_n^0 = \alpha_n$ is decoupled from all supplementary
fields.  Consider the commutator $[ B_m^j, W_n^0 ]$.  It only contains
operators $B_q^K, K < j$; hence once it is completely realized by the
representation with $j-1$ fields, the addition of a supplementary
residual field ${\bar W}_m^j$ does not modify the r.h.s., but it adds
a term $[ {\bar W}_m^j , \> W_n^0 ]$ to the left-hand side, hence one
must have $[ {\bar W}_m^j , \> W_n^0 ] = 0$.

We now study the algebraic properties of ${\bar W}_m^1$.  Consider
the commutator $[ B_m^j, \> B_n^1 ]$.  Once the $W_{1 + \infty}$
algebra is realized by $1+j$ operators, the addition of further
supplementary operators does not modify this commutator.  Since $[
B_m^j , \> B_n^1 ] \sim B^{(j)} + B^{(j-1)} + \dots$, the only term on
the r.h.s. containing ${\bar W}^{(j)}$ is $B^{(j)}$, according to
(2.10).  Finally, by assumption, this commutation relation is
satisfied by the operators obtained from the $j$-operator realization.
This leaves one single commutation relation to be satisfied:
$$
\left [ {\bar W}_m^j, \> {\bar W}_n^1 \right ] = (n-1)j \> {\bar
W}_{m+n-2}^j\eqno\eq
$$
Lower-order terms do not appear in this commutation relation.

Let us now construct the representation of $B_n^2$ and determine
the algebraic properties of ${\bar W}_m^2$.  Its
generic commutation relations with elements of the algebra follow from
the formula:
$$
\left [ B_m^j, \> B_n^2 \right ] = (nj - 2m)\> B_{m+n}^{j+1} + \dots
\eqno\eq
$$
First of all we determine the expression of $B_{m}^{j+1}$ when one
adds the supplementary $j$-spin field ${\bar W}_m^j$ to the $j$ fields
realization (conjectural) of $W_{1 +\infty}$.  Setting $B_m^{j+1} =
B_m^{j+1} (W^0 \dots \bar{W}^{j-1} ) + \Delta B_m^{j+1}
$ and inserting it in the commutator $[ B_m^{j+1},\> B_n^0 ]$ gives:
$$
\left [\Delta B_m^{j+1},\> W_n^0 \right ] = - n (j+1) {\bar
W}_m^j\eqno\eq
$$
The terms of $\Delta B_m^{j+1}$ containing $W_n^0$ are thus obtained
as:
$$
\Delta B_m^{j+1} = (j+1) \sum_\ell {\bar W}_{m-\ell}^j \> W_\ell^0 +
\dots \eqno\eq
$$
Dimensional considerations show that the remaining terms, which do not
contain $W_n^0$, are necessarily linear in ${\bar W}_m^j$ with a $m$-
dependent coefficient, since it must be $\sim \partial {\bar W}^j$.
{}From the commutation relation:
$$
\left [ B_m^{j+1},\> B_n^1\right ] = \left (n - m (j+1)\right )\>
B_{m+n}^{j+1} + m^2
j(j+1) \> B_{m+n}^j + \dots \eqno\eq
$$
and plugging into (2.22) the exact expression for $B_n^1$:
$$
B_n^1 = {1\over 2} \sum_\ell \> \colon\> W_{m-\ell}^0 \> W_\ell^0
\>\colon \> +\, {m-1\over 2}\, W_n^0 + {\bar W}_n^1\eqno\eq
$$
one finally gets:
$$
\Delta B_m^{j+1} = (j+1) \sum_\ell {\bar W}_{m-\ell}^j \> W_\ell^0 +
(j+1)\, {m-1\over 2}\,  {\bar W}_m^j \eqno\eq
$$
If we now add the supplementary field ${\bar W}^{j+1}$ to our
representation, $B_m^{\ell < j+1}$ is not modified, $B_m^{j+1}$ gets a
${\bar W}^{j+1}$ and $B_m^{j+2}$ acquires the additional term $\Delta
B_m^{j+2}$ obtained from (2.24).

The commutation relation of $B^{j+1}$ and $B^2$ now reads (from 2.10):
$$
\left [B_m^{j+1},\> B_n^2\right ] = \left\{n (j+1) - 2m\right\}
B_{m+n}^{j+2} + \left\{m^2 - n^2\> {j(j+1)\over 2} \right\}
B_{m+n}^{j+1} + \dots \eqno\eq
$$
The residual terms containing ${\bar W}^{j+1}$ obey therefore the
following algebra, once the recursive assumption that (2.10)
is realized by the
$j+1$ first fields, is taken into account:
$$\eqalign{
\left [ {\bar W}_m^{j+1} , \> B_n^2\right ] &= \left \{n (j+1) -
2m\right\} \left\{ (j+2) \sum {\bar W}_{m+n-\ell}^{j+1} \> W_\ell^0 +
(j+2)\, {m+n-1\over 2}\> {\bar W}_{m+n}^{j+1}\right\}\cr
&+ \left (m^2 - n^2\right ) \> {j(j+1)\over 2}\>\>
{\bar W}_{m+n}^{j+1}}\eqno\eq$$
An exact expression for $B_n^2$ is obtained from (2.15):
$$
B_n^2 = B_n^2 (W^0) + 2 \sum_\ell
{\bar W}_{m-\ell}^1 \>W_\ell^0 + (m-
1) {\bar W}_m^1 + {\bar W}_n^2 \eqno\eq
$$
Here $B^2 (W_n^0) \equiv B_n^2 (\alpha )$ is given by the standard
bosonization formula.  The second term describes an interaction
between the collective field $\alpha(z)$ and the spin--2 field
$\bar{w}_1 (z)$.

Substituting (2.27) into (2.26) finally leads to the remarkable
cancellation of all terms except $\left [ {\bar W}_m^{j+1} ,
\> {\bar
W}_n^2\right ]$.  It follows that the commutation relation (2.26),
now implemented on the $j+3$-fields representation, implies:
$$
\left[ {\bar W}_m^{j+1},\> {\bar W}_n^2\right ] = \left\{ n(j+1) -
2m\right\}\> {\bar W}_{m+n}^{j+2}\eqno\eq
$$
The cancellation of all lower order terms in (2.28) is a
non--trivial result, coming from the exact compensation in (2.25)
between the
r.h.s. residual contributions in ${\bar W}^{j+1}$ and the l.h.s.
contributions of the form $\left [ {\bar W}_m^{j+1},\> B_n^2 (W_0, \,
{\bar W}_1)\right ]$ using (2.27) and (2.18).

It now seems a reasonable conclusion that, if the recursive
construction can be consistently implemented at all orders in the
number of residual fields, the ``basic" residual fields ${\bar W}_m^j
(j\geq 1)$
obey a classical $w_\infty$ algebra.  The proof stated above is
complete for all commutators involving ${\bar W}^{(2)}$ and ${\bar
W}^{(1)}$.  The non--linear lower-order terms in $B_n^j$ allow the
elimination of lower-spin terms in the algebra of ${\bar W}_n^j$.
A similar construction for the
non--linear $W_N$ algebras, introduced in [19] and of which
$W_{1+\infty}$ is a particular
limit, can be viewed as a partial justification of
our scheme. It is detailed in Appendix A.

The higher--spin generators of $W_{1+\infty}$ can be computed
iteratively, although the explicit expressions become more involved.
We shall simply give the next non--trivial current.  It takes the
form:

$$\eqalign{
B_m^3 &= B_m^3 (W_0) + 6 \sum_\ell {\bar W}_{m-\ell}^1 \>B_\ell^1
(W_0) + 3 \sum_\ell\> (m-\ell) {\bar W}_{m-\ell}^1 \> W_\ell^0\cr
&+ \sum_\ell \>\> \colon {\bar W}_{m-\ell}^1 \> {\bar W}_\ell^1\>
\colon\>\>
+ c_m \>\> {\bar W}_m^1 + 3 \sum_\ell\> {\bar W}_{m-\ell}^2 \>
W_\ell^0\cr
&+ {3\over 2} (m-1) \> {\bar W}_m^2 + {\bar W}_m^3}\eqno\eq
$$
Here $B_m^3= B_m^3 (\alpha )$ is again the single boson representation of
$B_m^3$, obtained from standard bosonization;
$c_m$ is a $m$-dependent constant explicitly computable once the
normal-ordering $\colon\>\colon$ of the a priori ill-defined expression
$\sum {\bar W}^1 \> {\bar W}^1$ is consistently defined.

We have now
obtained a description of the bilinear fermion algebra -- which
contains all information about the spectrum of the matrix model -- in
term of a scalar field $w_0\equiv \alpha$, realizing a $U(1)$ current
algebra and an infinite decoupled set of higher spin fields $\{
\bar{W}_n^j \}$, realizing a centerless
$w_{\infty}$ algebra.  This description is given explicitely in (2.23,
27, 29).

The initial, pure $\alpha$ terms in the $W_{1+\infty}$ generators
are given by the
standard one--boson realization of the $W_{1+\infty}$ algebra.
In the semi--classical limit the $n$--spin generator of
$W_{1+\infty}$ is simply the $n$-th power of the current $\alpha$ (see
(2.14)).  Similarly, the $n$-th power of the (classical) collective
field $\alpha (x,t)$ realizes a $w_{1+\infty}$ algebra at the
classical level $\bigl(B_n^m \equiv \int {x^{m-1}\over m-n} \,
\alpha^{m-n}
dx\bigr)$, and even at the quantum level but without normal--ordering
[7].

To simplify notations, we now introduce one--index
.dimens of a continuous variable $z$
realizing the
centerless ``classical" $w_{\infty}$ algebra and encapsulating the
discrete 2--index generators $\bar{W}_n^j$.  For clear
spin--dimensional reasons, they are written as:
$$\bar{w}_{j} (z) = \sum \bar{W}_n^j \, z^{n-1-j}\eqno\eq$$
and describe the classical continuous version of $w_{\infty}$ found
for instance in [9].  Now it must be noticed that the representation
(2.9) of the $W_{1+\infty}$ algebra in terms of bilinears of free
fermions is not unique; there exists in fact an infinite discrete
set of isomorphic constructions defined as:
$$B_K^n (a) = \sum_{q\epsilon \IZ} : \psi^\dagger_{q-n - 1/2 + a}^
+ q^K \, \psi_{q-1/2 + a} \,\colon \, ; \, a\in\IZ \eqno\eq$$

Due to the equivariance of the contraction $\langle \psi_{q_1 -1/2
+a}^\dagger  \, \psi_{q_2 - 1/2 + a}\rangle \sim \delta_{q_1,q_2}$,
(2.31) obeys
the same algebraic relations (2.10) for any integer value of $a$.
Hence a given fermionic bilinear operator can be expressed in an
infinite (discrete) number of ways, depending on the choice of a
specific representation (2.31) of the bilinear algebra.

\section{Bosonization of the Hamiltonian}

Let us accordingly express the fermionic Hamiltonian -- which
summarizes the full dynamics of the theory -- in term of the
supplementary fields $\bar{w}_i (z)$ defined in (2.30) and the bosonic
field $w_0 (z)$ identified with the tachyon field $\alpha
(x,t)$.  We recall that the Hamiltonian (2.1) consists of a kinetic part
$H_K$ and a potential part $H_p$:
$$H_K = \int dz \, \partial\psi^\dagger \partial \psi \qquad H_p =
\int dz (U(z) - \mu_F ) \psi^\dagger \psi \eqno\eq$$

The potential part is the easiest to express; in fact it does not
depend on the choice of the shift $a$ in the definition (2.31)
since it only contains $K=0$ terms.  One has:
$$\int dz \, (U(z) - \mu_F) \psi^\dagger \psi = \int dz \, (U(z) -
\mu_F ) \sum_{r.r'} \, \psi_{-r}^\dagger \, \psi_{r'} \, z^{r+r' -
1}\eqno\eq$$
Each term $\sim g_b \, z^b$ in $U(z)$ generates a term:
$$U_b = g_b \, \sum_{r + r' = -b} \, \psi_{-r}^\dagger \psi_{r'} = g_b \,
\sum_{r'\in Z }\, \psi_{r' + b}^\dagger \psi_{r'} \equiv
 g_b \, B_{n=-b}^{K=0} \eqno\eq$$
whatever value $a$ takes in (2.31).

One immediately ends up, from (2.30) $(j=0)$, with:
$$H_p = \int \, U(z) \, \alpha (z) \, dz\eqno\eq$$
The kinetic part is more tricky.  From (2.32) one gets:

$$\eqalign{ H_K & = \sum_{r,r'} \, \psi_{-r}^\dagger \, \psi_{r'} \, (r-
1/2) (r'-1/2) \oint z^{r+r'-3} dz\cr
& = - \sum_{r'} \, \psi_{r'-2}^\dagger \, \psi_{r'} \, (r'-1/2)
(r'-3/2) }\eqno\eq$$
and now the expression of $H_K$ in terms of $B$--variables depends
explicitely on the choice of the shift--variable $a$ in (2.31):
$$H_K = \, B_{n=2}^{K=2} + (2a -3) \, B_{n=2}^{K=1} + (a^2 - 4a +
9/4) \, B_{n=2}^{K=0} \eqno\eq$$

The choice of a suitable value for $a$ will be determined by
considering the expression
of (2.37) as a function of the fields $w_0 (z) \equiv \alpha (z)$
and $\bar{w}_1 , \bar{w}_2$.  From (2.37) and (2.24) we get:
$$\eqalign{ H_K & = B_2^2 (\alpha ) + 2 \sum_l \, \bar{W}_{2-l}^1 \,
\alpha_{\ell} + (2a - 2) \bar{W}_2^1 + \bar{W}_2^2\cr
& + {(2a-3)\over 2} \sum_l \, \alpha_{2-\ell} \, \alpha_{\ell} +
{2a-3\over 2} \,
\alpha_2 \, + (a^2 - 4a + 9/4 ) \alpha_2\cr
& = P (\alpha) +  2 \int dz \left( \bar{w}_1 (z) \alpha
(z)\right) + (2a-2)
\int dz \cdot z^{-1} \bar{w}_1 (z)\cr
& +  \int dx \bar{w}_2 (z)\, . }\eqno\eq$$

We expect to get a Hamiltonian which reduces to a pure cubic
interaction term when the sole field $\alpha$ is kept; hence
$P(\alpha )$ must be identified with ${\alpha^3\over 3}$.  Moreover,
since the initial Hamiltonian (2.32) is
translation--invariant, the supplementary terms in (2.38) also have to
be translation--invariant.  Translation--invariance
is only achieved when $2a-2=0$ or $a=1$.  The
resulting complete Hamiltonian describing the tachyon $+$ discrete
fields and reducing to the pure tachyonic form (1.2) when
$\bar{W}\equiv 0$ is thus conjectured to be:
$$H = \int \left\{ {\alpha^3 (z)\over 3} + (U(z)-\mu_F ) \alpha(z) +
\bar{w}_1
(z) \alpha (z) + \bar{w}_2 (z)\right\} dz \vert_-^+\eqno\eq$$
and the dynamics of the classical theory is described by the Poisson
structure:
$$\eqalign{ \{ \alpha (z), \alpha (z') \} & = \delta ' (z-z' ) ;
\qquad \{ \alpha (z), \bar{w}_i (z) \} = 0\cr
\{ \bar{w}_i (z) , \bar{w}_j (z') \} & = (i \bar{w}_{i+j-1} (z) + j
\bar{w}_{i+j-1} (z') ) \delta ' (z-z') }\eqno\eq$$

One has in fact two chiralities $\alpha_{\pm}$ corresponding to
the two independent and decoupled sets of generators $w_0^n (\pm
)$ associated with the two independent sets of non--relativistic
chiral fermions describing the matrix model.
The higher--spin fields $\bar{W} (j)$
also realize two copies $\bar{W}_+^j$ and $\bar{W}_-^j$ of
$w_{\infty}$, corresponding to the original two sets of chiral Dirac
fermions (2.6,7).  Their full physical relevance will be commented on
later.

This improved formulation leads to the following sequence of coupled
equations of motion for the fields $w_0, \bar{w}_j$.
$$\eqalign{ \partial_t \bar{w}_0 & = \dot{\alpha} = \{ H,\alpha\} =
\alpha \partial_x \alpha + \partial_x U + \partial_x \bar{w}_1\cr
\partial_t {\bar{w}}_1 & = \{ H, \bar{w}_1\} = -2 \partial_x \alpha
\cdot \bar{w}_1 -
\partial_x \bar{w}_1 \cdot \alpha + \partial_x \bar{w}_2\cr
\partial_t {\bar{w}}_2 & = \{ H, \bar{w}_2 \} = -3 \partial_x \alpha \cdot
\bar{w}_2 - \partial_x \bar{w}_2 \cdot \alpha + 2 \partial_x
\bar{w}_3\cdots\cr
\partial_t {\bar{w}}_j & = \{ H, \bar{w}_j \} = - (j-1)
\partial_x \alpha \cdot \bar{w}_j -
\partial_x \bar{w}_j \cdot \alpha + j \partial_x \bar{w}_{j+1}\cdots}
\eqno\eq$$

We now summarize the final result of this section which is essentially
the main result of the paper.  We have seen that the quantum dynamics
of 2--dimensional fermions can be fully described in terms of a
bosonic (collective) field and an infinite sequence of discrete (later
characterized as topological) fields closing a $w_{\infty}$ algebra.
The latter are constrained to have their central charge set to zero.
The two distinct sets of fields enter the complete Hamiltonian and
interact through a direct coupling, uniquely determined by the
bosonization scheme.
\endpage

{\bf\chapter{\bf Realizations of the $w_{\infty}$ Algebra}}

We have in the previous section introduced an abstract algebra of
higher spin fields
realizing (classically and quantum--mechanically) a $w_{\infty}$
centerless structure.  We now wish to explore this structure in order
to understand the physical meaning of this algebra.  We
shall address two issues:  the possibilities of reduction to a finite
number of -- still abstract -- fields, and the representation of the
full algebra $w_{\infty}$ by explicit fields.  This will lead us to a
three--dimensional structure related to the large--$N$ KdV
hierarchies.

\section{Truncations of the Field Content}

Because $\{\bar{w}_j (z) \}$ realize a centerless
$w_{\infty}$ algebra, it is consistent to truncate this algebra at any
level and
some of these truncations have a particular physical interest.

First of all, one can (trivially) choose to set all ``discrete" fields
${\bar w}_j (z)$ to 0, which leaves us with the one--boson realization
and correspond to the
representation of the $W_{1+\infty}$ (fermionic) algebra by the normal
ordered polynomials in the collective (tachyon) field $\alpha_{\pm}
(z)$.

A more interesting choice is to keep one single supplementary field
$\bar{w}_1 (z)$,
describing a $c=0$ ``energy--momentum tensor" for some implicite
underlying
topological field theory. This notion will very soon be further
developed, when we shall examine realizations of the $w_{\infty}$
algebra.

There actually exists in the literature [20] a 2--boson
representation.  The two constructions are in fact
similar but not actually related.
For one thing, the two--boson realization cannot be reduced to
a one--boson realization by eliminating the
second field, because OPE of the second field give back the first
field!  In this two--boson representation, one has:
$$\eqalign{ w_0 & = \partial\sigma\cr
w_1 & = - {1\over 2} (\partial\sigma )^2 + \bar{w}_1 \, ; \, \bar{w}_1
\equiv {1\over 2} \left ( (\partial \rho )^2 - \partial^2 \rho\right)\cr
w_2 & = {1\over 3} (\partial\sigma )^3 + 2 (\partial\sigma ) \bar{w}_1
+ \bar{w}_2 ;\cr
& \bar{w}_2 \equiv {1\over 3} \partial \bar{w}_1 - {4\over 3}
(\partial \rho) \bar{w}_1\cr
w_3 & = -{1\over 4} (\partial\sigma )^4 - {1\over 10} (\partial\sigma
\,\partial^3 \sigma ) + {3\over 20} (\partial^2 \sigma )^2\cr
& - 3 \, \partial\sigma \cdot \bar{w}_2 + 3 (\partial \sigma )^2
\bar{w}_1 + \bar{w}_3 ;\cr
& \bar{w}_3 = {4\over 3} (\bar{w}_1 )^2 +\cdots }\eqno\eq$$

One sees in (3.1) the same kind of structure as in the reexpressions
of the fermionic generators $B_n^m$ in (2.27) and (2.29), namely a
purely $(w_0 \equiv \partial\sigma )$--dependant term, plus
similarly--looking coupling terms involving
$\partial\sigma$ and particular $\partial \rho$--dependant
currents $\bar{w}_j (x)$, plus the pure ``supplementary" currents.
The difference however is that the constituent field $(\partial
\rho)$, not simply its derived composite field $\bar{w}_1 (z)$,
appears explicitely in the expressions of higher currents as function
of lower currents, which is not the case in our construction, nor in
the nested construction [19]
of $W_N$ as $(j_0 \oplus W_{N-1}$) described in Appendix A, nor in the
related construction of $W_3$ as a 2--field algebra [21].
The reason for this difference is
mentioned above:  the 2--boson construction is ``intrinsically" a
2--field construction, and certainly not an ``additive" or nested
construction adding one field to another.  The similarity of
structure merely reflects the
fact that in both cases, the fixed spin values of the currents put
constraints on the allowed terms.

Our representation of $W_{1+\infty}$ by these two (tachyon +
centerless energy--momentum tensor) fields is actually an underlying
feature of the perturbative
computations for the collective field theory described in [4].
The non--local vertex which arises there can actually be replaced by a
propagator involving a $\bar{w}_1$--like supplementary field.
Furtherly ``truncated" theories may have similar physical
applications, which we will discuss here.

Finally one can decide to keep all $w_{\infty}$--algebra fields.  We
now turn to the problem of representing the $w_{\infty}$ algebra in a
more self--contained way, and we shall introduce
two seemingly different representations of this algebra, both using a
single, 3--dimensional field as a basic ingredient.

\section{The 2+1--Toda Current Representation}

The first construction uses the Toda 2+1 chiral current
representation.  As a starting point, we recall that the
large--$N$ limit of the classical non--linear $W_N$
algebra [22] can be
consistently defined as a classical linear $w_{\infty}$--algebra.
One starts from the Gelfand--Dikii
Poisson structure of the $N$th KdV hierarchy (which is a classical
$W_N$ algebra [23]) and defines a
suitable rescaled large--$N$ limit [9].  There exists also a realization
of a $w_{\infty}$--algebra which takes more direct advantage of the
relation between the $sl(n)$ Toda theory and the $WA_n$ algebras
[23,19], and
considers the large--$N$ limit of the $A_N$ Toda hierarchy to be directly
constructed as a $2+1$ dimensional theory [16,24] instead of being
obtained by a limit procedure [25] from a $WZW$ model.

In any case, the chiral currents $w_j (z)$ of the 2+1 Toda theory
realize classically a $w_{\infty}$ Poisson algebra [9].  This
construction was
explicitly realized in [26] using an approach based on the Lax
representation of the theory as a 2-dimensional $w_\infty$ Toda field
theory [24,16].

The $2+1$ dimensional Toda theory is described by the following
equation of motion:
$$
\partial {\bar \partial} \> \phi (z, {\bar z}, s) = - \exp
\partial_s^2 \phi\eqno\eq
$$
or, defining $\partial_s \phi \equiv u\,\colon\,\partial {\bar
\partial} u = - \partial_s \exp \partial_s u$.  The Lax representation
of (3.2) uses the functional representation of the
classical $w_\infty$ algebra as a Poisson algebra of functions of two
conjugate variables:
$$
w_\infty \simeq \left\{ F(\lambda, s) \> ;\> [F, G]_{Lie} = \lambda
\left ({\partial F\over {\partial \lambda}} {\partial G\over {\partial
s}} - {\partial F\over {\partial s}} {\partial G\over {\partial
\lambda}}\right )\right\}\eqno\eq$$
It reads:
$$
\eqalign{
L &\equiv \lambda + \mu_0 (s) + \sum_{n \geq 1} \lambda^{-n} u_n
(s)\cr
{\hat L} &\equiv \sum_{n \geq 1} \lambda^n {\hat u}_n
(s)\cr
{\bar B}_{(n)} &\equiv ({\hat L})_{\leq - 1}^{-n}\> \left ({\rm
projection\>of\>} ({\hat L})^{-n} \> {\rm on}\> \left\{ \lambda^p, \> p
\leq - 1\right\} \right )\cr
B_{(n)} &\equiv (L)_{\geq 0}^n \>\> (\rm same\> notation)\cr
{\bar \partial}_{(n)} L &= [ {\bar B}_{(n)}, L ]_{Lie} \qquad\qquad
\partial_{(n)} L = [ B_{(n)}, L ]_{Lie}}\eqno\eq$$
(3.4) describes the full 3-dimensional Toda hierarchy with
respect to the two infinite sets $(z_n, {\bar z}_n \dots)$ of light-cone
variables.  It appears as a very natural generalization of the
Lie-algebra Toda theories in 2-dimensions [27] ((3.2) follows from
(3.4) with
$z_1 = z, \>{\bar z}_1 = {\bar z}, \> \mu_0 = \partial \partial_s
\phi$).  (3.2) can also be obtained by using the construction of
$w_{\infty}$ as a continuum $n \rightarrow \infty$ limit of $s\ell (n,
{\bf C})$ [9,16].

The chiral currents $w_n$ (${\bar \partial} w_n = 0$ as a consequence
of (3.2)) are obtained from natural candidates for chiral
densities, namely the ``adjoint-invariant" Adler traces [28]
$\ointop d\lambda
{ds\over s} L^n$.  Using (3.4) to redefine such expressions only in terms
of $\phi$ allows to get the first chiral currents realizing the
$w_{\infty}$ algebra, denoted $w_2, w_3$ and $w_4$; in principal,
the Poisson bracket algebra then gives the complete set of currents
simply from $w_2$ and $w_3$, although no closed form expression is yet
available beyond $w_5$.  It is in fact a safe conjecture based on the
form of the Adler traces from (3.4), that non--local densities will
arise when trying to construct consistent higher--spin generators
[29].
One has:
$$
\eqalign{
w_1 & = \int ds \left ( {(\partial \partial_s \phi )^2\over {2}} -
\partial^2 \phi\right )\cr
w_2 &= \int ds \left( \>\> {-(\partial \partial_s \phi )^3
\over {3}} + 2 (\partial
\partial_s \phi ) \partial^2 \phi + s (\partial^3 \phi + \partial^2
\phi \> \partial \partial_s^2 \phi )\right)}\eqno\eq
$$
with the Poisson bracket structure:
$$
\left\{ \partial \partial_s \phi (s, z),\,\, \partial
\partial_s^\prime \, \phi (s^\prime, z^\prime\, ) \right\} =
\delta^\prime \, (z - z^\prime\, )\> \delta (s - s^\prime\,
)\eqno\eq
$$
(considering $\bar z$ as a time-variable which therefore does not
appear in the Poisson bracket.)  This set now generates the full
algebra of chiral currents.

We show in Appendix C that (3.5) can
be obtained directly from continuous limits of the $w_n$  chiral
currents obtained for $W_N$ algebras in the Feigin-Fuchs construction
[30].

{}From this
realization follows an explicit expression for the matrix model
Hamiltonian, describing the coupling between the two--dimensional
tachyon field $\alpha (x,t)$ and the 3--dimensional field $\phi
(z,\bar{z}, s)$:
$$\eqalign{ H & = \int \left\{ {\alpha^3(z)\over 3} + (U(z) -
\mu_F ) \alpha (z) \right\} dz + \int  2\left\{ \alpha (z) \cdot
\left ( {(\partial\partial_s \phi )^2\over 2} - \partial^2 \phi \right
) \right\} dz \, ds\cr
& + \int dz \, ds \, {-(\partial\partial_s \phi )^3\over 3} +
\cdots }\eqno\eq$$
with the Poisson bracket structure:
$$\eqalign { \left\{ \alpha (z) , \alpha (z') \right\} & =
\partial_{z} \, \delta (z-z') \, ; \, \left\{ \partial\partial_s
\phi (z,s), \partial_s \phi (z', s' )\right\} = \delta (z-z') \delta
(s s')\cr
& \left\{ \phi (z,s), \alpha (z')\right\} = 0 } \eqno\eq$$

Some comments about this particular Hamiltonian realization are
relevant at this point.  First of all, it has the interesting feature,
recalling the constructions in [12], of encapsulating the interaction
of the whole set of fields $\{ \bar{w}_j (z)\}$ as a single
interaction term involving a 3--dimensional field.  It has a uniquely
determined coupling to the tachyon, and a $c=0$ central charge.
Notice that the leading--order--power terms in (3.7) can actually
be recast
as a pure cubic interaction of a composite field $(\alpha -
\partial\partial_s \phi)$, namely:
$$H = \int dz \, ds \, {(\alpha - \partial\partial s\phi )\over 3}^3 +
(\alpha - \partial\partial_s \phi ) \, \partial^2 \phi + \cdots
\eqno\eq$$

This statement is correct provided the Adler convention $\oint ds
\equiv$ Residue $(s^{-1})$ is not applied to the pure $\alpha$--term,
but simply keeps it as an $s$--independent term ``beyond" the algebra
$Ps$Diff.  This suggests an interpretation of (3.7) as arising from a
non--semi--simple $Gl(\infty)$--type Toda system, with non--trivial
boundary conditions described by the 2--dimensional field $\alpha(z)$.
However the issue is made obscure by the occurence in (3.9) of
subleading terms $\sim (\alpha - \partial\partial_s \phi)
\partial^2\phi$,
which cannot obviously be recast purely in terms of $(\alpha -
\partial\partial_s \phi)$.

Concerning the relevance of these terms, one must recall that the
realization (3.5) - (3.6) of $w_{\infty}$ is specifically a
construction of currents constrained by the chirality condition
$\bar{\partial} w = 0$, realized through the explicit Toda equation
(3.2).  Hence it may appear irrelevant in our framework to consider
this particular set of currents instead of the simpler, non--chiral
objects $\int {(\partial\partial_s \phi )^n\over n} \, ds$, since we
only need some realization of $w_{\infty}$.  Our choice can however be
justified on the following grounds.

It is known [31]
that the quantized algebra of chiral currents for $N\rightarrow\infty$
KdV hierarchies (which is directly related to
the 2+1 Toda algebra [35,9]) is a {\it classical} $w_{\infty}$--algebra,
with a sole
central charge in the Virasoro sector.  It might then be sufficient to
define an ``improved" quantum $\bar{w}_1(z)$ current, or
(in order to assure a centerless algebra) couple ghost--fields to the Toda
theory. This argument shows that the Toda chiral
invariant densities should be understood in fact as ``improved"
$w_{\infty}$ densities realizing a classical -- except in the Virasoro
sector -- algebra after quantization.  They are therefore a
natural choice in our framework, since we are ultimately interested
in realizing a
{\bf quantum} $w_{\infty}$ algebra.

\section{The Topological Representation}

There exists a construction of the $w_{\infty}$ currents
where the centerless condition arises naturally.  Indeed, a
more obvious way to realize a $c=0$ $w_{\infty}$
algebra--which  must
have a spectrum consisting only of ``global", or topological, discrete
excitations -- is to use explicitely topological densities, i.e. total
derivatives of some fields with a suitable Poisson structure.
Such densities can be given by the conserved Hamiltonian
densities of the KP
hierarchy, once they are expressed in terms of coefficients of
the Baker--function
[28].  The Baker function
coefficients are in fact coordinates on the coadjoint orbit of
$Ps$Diff [32] describing not the algebra element itself, but the group
element whose coadjoint action on a given initial point leads to the
algebra
element.

To be more specific, and see how the topological nature of those
densities arises, let us recall the form of the first few Hamiltonian
densities for the KP hierarchy, for instance found in [33].

Defining the KP Lax operator in $Ps$Diff as:

$${\cal L} = \partial + u_1 \partial^{-1} + u_2 \partial^{-2} +
\cdots\eqno\eq$$
the coordinates on the ``orbit" are the functions $u_i (x)$, and the
(first KP) Poisson structure is the Kirillov bracket for the algebra
$Ps$Diff.
In fact, (3.10) such as it stands does not represent a single orbit of
the $Ps$Diff algebra, but the full algebra $Ps$Diff$_-$, and the
Kirillov bracket is accordingly degenerate.
The Baker function $K(x,\partial )$ is defined so that ${\cal L}$ can
be rewritten as:

$$\eqalign{ {\cal L} = & K^{-1} (x,\partial )\cdot \partial \cdot K
(x,\partial )\cr
K (x,\lambda )& \equiv 1 + \sum_{j=1}^{\infty} a_j \lambda^{-j}
}\eqno\eq$$
$\lambda$ being a formal second variable for $K$ (a sort of spectral
parameter associated with $x$). $K$ defines
a second set of coordinates $a_j$, which can indeed be
understood as more ``group" than ``algebra" -- like objects.

They are
related to the original coordinates by:

$$\eqalign{ u_1 & = - \partial_x a_1\cr
u_2 & = - \partial_x a_2 + a_1 \partial_x a_1\cr
u_3 & = - \partial_x a_3 + (a_1 a_2 ), x - a_1^2 a_{1,x} -
a_{1,x}^2\cr
u_4 & = - \partial_x a_4 + \cdots }\eqno\eq$$

On the KP ``orbit" (3.10) exists an infinite set of conserved
Hamiltonians, for instance computed by expanding the eigenfunction
$\psi$ of
${\cal L}$ in (3.11) in inverse powers of the eigenvalue $E$ and more
fundamentally obtained as Adler traces [28,34] $\Tr\, {\cal L}^n
\equiv$ $\int {\rm Residue} \, \vert_{\partial^{-1}} {\cal L}^n
\cdot dx$.  The corresponding
Hamiltonian densities read:
$$\eqalign {w_1 &  = - u_1\cr
w_2 & = u_2\cr
w_3 & = -u_3 - u_1^2\cr
& \cdots }\eqno\eq$$
and can therefore be rewritten as total derivatives in terms of the
Baker function coefficients:

$$\eqalign{ w_1 & =  - \partial_x a_1\cr
w_2 & = \partial_x \left ( - a_2 + {a_1 ^2\over 2} \right ) \cr
w_3 & = \partial_x \left( - a_3 + a_1 a_2 - {a_1 ^3\over 3}\right)
}\eqno\eq$$

The form of the Hamiltonian densities as total derivatives of the
Baker function coefficients follows from general properties of the
coadjoint orbits [34].
We must now realize a $w_{\infty}$--algebra from
the Poisson structure of these Hamiltonian
densities.  It is known [9] that the Hamiltonian densities for the
long wavelength limit of KP indeed realize a $w_{\infty}$ algebra,
once they are rewritten as functionals of the coefficients, when
$n\rightarrow \infty$, of a $n$--th order KdV Lax operator with a
2nd Gelfand--Dikii structure [35] for its coefficients.  Details of the
demonstration are given in Appendix B.

Defining the derived KP operator (3.10) from a large $N$ KdV operator
as:$$\hat{{\cal L}}_{\infty} = \lim_{N\rightarrow\infty} \, \left [
\partial^N + v_{N-2} \partial^{N-2} + \cdots + v_0 \right
]^{1/N}\eqno\eq$$
the conserved hamiltonian densities are canonically [33]:
$$ h^{(n)} = {\rm Res} \,\, \left( {\cal L}^{n/N} \equiv
\hat{{\cal L}}_{\infty}^n \right) = {\rm Res} \, K \partial^n K^{-
1}\eqno\eq$$
This gives therefore an expression of the form (3.14) for the
Hamiltonian densities. One must consider the long wave--length
limit, i.e. all
derivatives of a function beyond first order are cancelled.  The
equations (3.12-14) are accordingly simplified (although only first
derivatives appear in the first 3 Hamiltonians (3.14)), but the
Hamiltonian densities generalizing (3.14) remain total derivatives.
The
matrix model Hamiltonian reads, in term of this representation:
$$\eqalign{ H & = \int dz \left\{ {\alpha^3\over 3} + (U (x) -
\mu_F ) \alpha (z) + \alpha(z) \,\,{\rm Res} \, (K\partial^2 K^{-1}
)\right\}\cr
& + {\rm Tr}\,\left( K \partial^3 K^{-1}\right)\cr
{\rm where} \,\, K & \equiv 1 + \sum_{j=1}^{\infty} \, a_j
\partial^{-j}\, ; } \eqno\eq$$
Again one has a representation of the full set of fields
$\bar{w}_j(x)$ using one single, 3--dimensional ``field" $K
(z,\partial,t)$.

\section{The Large-$N$ KdV Framework}

These two constructions actually correspond to two different
parametrizations of the large $N$ limit of the
$N$--th order KdV Lax operator, in the long wavelength limit.  Their
connection to this formalism, and the subsequent relation between
them, can better be summarized in the following diagram.

\item{(1.b)} \underbar{$N\rightarrow \infty$ continuous limit}
[16]
\itemitem{$\bullet$} $\partial_z \phi
(z,s) \equiv \phi_n (z)$.  Field
variable
is a 2+1 Toda field; Poisson structure is a U(1) current
algebra.
\itemitem{$\bullet$} $H^{(n)} = \oint
ds \, P(\partial_s \phi )$ (eqns.
3.5,[9.25]).  Hamiltonian densities realize a $w_{\infty}$ algebra.
(long wavelength limit)

$${\bf{\updownarrow}}$$

\item{(1.a)}  \underbar{Miura--
transformed, discrete theory }
[23,34]
\itemitem{$\bullet$}  ${\cal L} = \prod_{n=1}^N \, (\partial +
\phi_n )$
\itemitem{$\bullet$}  $\phi_n (z) \equiv \phi_n (v (z))$.  Field
variables are Miura--transformed of the KdV potentials, obeying a
$sl(n,{\bf C} )$
Kirillov--Poisson algebra, or Toda $sl(n,{\bf C} )$--fields.

$${\bf{\updownarrow}}$$
\underbar{Initial model}:  \underbar{$N-th$ order KdV} [34]
\itemitem{$\bullet$} ${\cal L} = \partial^N + v_2 \partial^{N-2} +
\cdots v_N$\itemitem{$\bullet$} $\{ v_j (z)\}$:  Field variables
are KdV potentials, obeying
second Gel'fand-Dikii Poisson structure [34].
\itemitem{$\bullet$} $H^{(n)}$ = $H^{(n)} (v_j ) $ = Residue
$(\partial^{-1}) {\cal L}^{n/N}$ [28,34]
\itemitem{$\bullet$} By definition ${\cal L}^{1/N} $= $\partial +
u_2 \partial^{-
1} + \cdots u_j \, \partial^{-j+1} + \cdots ; u \equiv u (v_j )$

$${\bf{\updownarrow}}$$

\item{(2a)}  \underbar{$N\rightarrow \infty $ discrete limit}
[9]\nextline
+ long wavelength limit for $u(v_j)$ and Poisson structures.
\itemitem{$\bullet$}  $\hat{{\cal L}} = \partial + \, u_2 \,
\partial^{-1} \cdots
u_j \partial^{-j+1} + \cdots$
\itemitem{$\bullet$}  $\tilde{u} \equiv \sum_{j\geq 2} u_j
\lambda^{-j} \, ; \,
\tilde{v} \equiv \sum_{j\geq 2} \, v_j \, \lambda^{-j} \Rightarrow
\tilde{u} = \ln (1 + \tilde{v})$
\itemitem{$\bullet$}  Field variables are $\tilde{u}$, obeying a
derived Poisson algebra.
\itemitem{$\bullet$}  Hamiltonian densities $H^{(n)} \equiv u_{n+1}$ when
$N\rightarrow \infty$.  They realize an exact $w_{\infty}$--current
algebra.

$${\bf{\updownarrow}}$$

\item{(2b)}  \underbar{Baker function parametrization} [28]
\itemitem{$\bullet$}  $\hat{{\cal L}} = K\cdot\partial\cdot K^{-1} \,
; \quad K
\equiv 1 + \sum_{s\leq 1} \, a_s \partial^{-s}$; long wavelength
limit.
\itemitem{$\bullet$} Field variables are group--like coordinates on
$Ps$Diff, with derived Poisson structure; $a_s \equiv a_s (v)$.
\itemitem{$\bullet$}  $H^{(n)} \equiv H^{(n)} (a)$ are
$w_{\infty}$--current
densities {\bf and} total derivatives in term of $a$--fields.

Labelings 1 and 2 refer respectively to the two constructions 3.2 and
3.3 previously
described.  These two representations of a $w_{\infty}$ classical
current algebra are in this way related.  It may however be more
interesting to consider either one or the other specifically, for
particular purposes.  We are now going to describe an interesting
algebraic analogy between our model and the well--known 2+1
topological gauge field theory with a Chern--Simons term, which
indicates that the purely topological representation has
possibly more relevance, or a deeper meaning.

\section{Analogy with Chern--Simons TGT}

The Hamiltonian (3.17), involving a 2--dimensional continuous degree of
freedom $\alpha (z)$, plus ``discrete" topological degrees of
freedom  $\bar{w}_j (z)$, calls to mind the
structure of Hamiltonians for 2+1 topological $U(1)$ gauge theories
involving a  Chern--Simons topological
term [36].  This analogy was already noticed in [37];  the TGT
contains indeed a  decoupled
lagrangian for a discrete, ``topological" or global degree of freedom
for the gauge field.  The corresponding Hamiltonian belongs to one of
two $w_{\infty}$--algebras which can be constructed out of the global
gauge degree of freedom; more specifically it is the $W_1^0$ component
of this algebra $\sim a^+ a + aa^+, a,a^+$ being creation/annihilation
operators for an effective harmonic oscillator.  In our fermionic
language the corresponding Hamiltonian is given by the momentum operator:
$$P = \int  \psi^{\dagger} \partial \psi \, dz\eqno\eq$$
{}From a similar analysis as for the kinetic
Hamiltonian (2.32), one gets essentially:
$$P = B_{n=1}^{K=1} \eqno\eq$$
which becomes, once expressed in terms of the completed set of bosonic
fields.
$$P = \int P(w_0) dz + \int \bar{w}_1 (z) dz\eqno\eq$$

This leads to a theory of a massless 2--dimensional
free particle plus
a decoupled degree of freedom with a Hamiltonian realizing the $W_1^0$
component of a $w_{\infty}$--algebra, exactly similar to the
Hamiltonian for the gauge degree of freedom of the TGT.

This nice algebraic analogy seems to indicate that an
explicit ``topological" representation of the supplementary fields
$\bar{w}_j (z)$ given in terms of total derivatives has some relation with
a deeper underlying gauge theory, and in this sense is the most
relevant parametrization.
We shall from now on choose this representation, and even when no
explicit relization is used, keep in mind the ``global", or
total--derivative, nature of the fields $\bar{w}_j (z)$.
It is in this quantization of additional
discrete fields that our approach differs from the others.  Usually
one would think of treating all the fields equally.
\endpage

{\bf\chapter{Relevance of the Scheme}}

We have derived in the previous sections an extension of the bosonic
collective field theory.  In addition to the scalar collective field we
now have an infinite sequence of higher spin fields $\{w_n (x,t)\}$
with a well defined interaction (2.39) between the two sets of degrees of
freedom.  As it was emphasized, we have found it useful to
characterize this space--time fields with a central charge.  While the
scalar boson carries the full central charge of $c_{\alpha} = 1$ the
additional $w_{\infty}$ fields have $c_w = 0$.  This implies a strong
constraint on the dynamics and quantization of these additional fields
which in our framework are
seen to be topological.  Their self--dynamics is given by hamiltonian
densities taking the form of total derivatives (3.14,17).  In
quantization one is to supplement these fields with appropriate ghost
fields whose presence is to assure that the central charge is kept to
zero.  We stress that we have introduced this notion of central charge
for target space fields as compared with standard world sheet conformal
field theory charges.

Even though the additional topological fields have essentially no
dynamics, they do couple in a specific way
to the collective boson.  They are then of importance in the
combined system.  In the remainder of this section we will point to
some of their basic physical properties and relevance.

Consider first the ground state in the presence of some potential
$v(x)$ (we think for instance of the inverted oscillator $v (x) = -x^2$
relevant for string theory).  The equation of motion for
$\alpha$ in the static case becomes
$$ \left( \alpha_{\pm} (x)^2 - x^2 + \mu + \bar{w}_1^{\pm}
(x) \right) = 0\eqno\eq$$
$$-2 \alpha_{,x} w_1 - \alpha \partial_x w_1 + 2\partial_x w_2 = 0$$
These equations are now valid for {\bf all} $x$ as compared with the
single boson collective equation
$$\partial_x \left( \alpha_{\pm}^2 - x^2 + \mu \right) = 0 \quad ,
\quad \vert x \vert \geq x_0 = \sqrt{\mu}\eqno\eq$$
which was defined on the limited domain (outside of turning points).
As before we have the static
collective field
$$\alpha_{\pm}^0 =
\cases{ \pm \sqrt{x^2 - \mu} & $\vert x \vert \geq
\sqrt{\mu}$\cr
0 & $\vert x \vert \leq \sqrt{\mu}$\cr}\eqno\eq$$
which vanishes inside the potential.  The additional spin two
field is now
$$w_1^0 = \cases {0 & $\vert x\vert \geq \sqrt{\mu}$\cr
x^2 - \mu & $\vert x \vert \leq \sqrt{\mu}$\cr }\eqno\eq$$
and it is nonzero under the potential.  The result is that the
variational equation holds for all $x$ (everywhere in space).  One
notes at this point that if one represented the spin 2 field $\bar{w}_1
(x)$ by a scalar field $\bar{w}_1 (x) = \varphi (x)^2$ the ground state
would read.
$$\varphi_0 = i \sqrt{\mu - x^2} \quad  x \in [-x_0 , x_0]
\eqno\eq$$
with an imaginary value.

Consider next the small fluctuations:  $\alpha = \alpha^0 + \eta,
\bar{w}_1 =
w_1^0 + \tilde{w}_1$.  The collective scalar field is defined on the
noncompact interval outside the potential $x\in [ -\wedge , -
x_0 ], x\in [ x_0, \wedge ]$ with the physical length equaling
$$L = \int_{\wedge}^{x_0} \, {dx\over \pi \phi_0 (x)} \rightarrow
\infty\eqno\eq$$
$\wedge$ being some large, positive cutoff.  The supplementary field
$\tilde{w}_1$ has the commutation
$$\left[ \tilde{w}_2 (x) , \tilde{w}_2 (y) \right] = - 2 w^0(x)
\partial \delta (x-y) + \partial w^0 \delta (x-y) \eqno\eq$$
and is consequently defined on a {\bf compact} domain with
$$iT = \int_{-x_0}^{x_0} {dx\over \sqrt{w_0 (x) }} = i \int_{-
x_0}^{x_0} \, {dx\over \sqrt{\mu - x^2}}\eqno\eq$$
The physical picture is that of a collective scalar boson defined on
the noncompact domain with a continuous spectrum and
a sequence of discrete imaginary energy fields defined on the compact
region $(-{T\over 2} , {T\over 2} )$ under the potential (tunneling
region).  This is depicted in Fig. 1. It is a general feature of this theory
that one finds the collective bosonic fields defined on the
classically allowed (noncompact) domain while the discrete fields
are defined on the classically forbidden compact interval.

The two sets of degrees of freedom are not decoupled. They
couple precisely at the boundary $(x = \pm x_0)$ as we now show.
Consider the small fluctuations equations of motion.  In standard
collective field theory one had $\pi\phi = \alpha_+ - \alpha_-$ with
$N = \int dx \phi$ being a conserved quantity.  Consequently
$$0= \dot{N} = \int_{\wedge}^{x_0} \,dx \dot{\phi} = \int_{
\wedge}^{x_0} \ \, \partial_{x'} \left(
\alpha_+^2 - \alpha_-^2\right)\eqno\eq$$
which implied that at the boundary
$$\alpha_+^2 (x_0 , t) = \alpha_-^2 (x_0 , t)\eqno\eq$$
For small fluctuations
$$\alpha_{\pm} = \pm \pi \phi_0 + {1\over \pi \phi_0}
\tilde{\epsilon}_{\pm} \eqno\eq$$
this translates into
$$\tilde{\epsilon}_+ (t, x_0 ) = - \tilde{\epsilon}_- (t, x_0 )
\eqno\eq$$
which are Dirichlet boundary conditions.  In the mode expansion
$$\epsilon_{\pm} (t,\tau) = \int dp \, e^{-ip(t\pm \tau)}
a_{\pm} (p)\eqno\eq$$
One only has {\bf half} of the possible degrees of freedom and
$$a_+ (p) = - a_- (p) = a(p)\eqno\eq$$
and
$$a_+ (p) + a_- (p) = 0\eqno\eq$$
by Dirichlet boundary conditions.

In our improved classical equation we now also have the spin two
$\bar{w}_1$
field,
$$\dot{\alpha} = \partial_x \left( \alpha^2 - x^2 + \mu +
\bar{w}_1\right)\eqno\eq$$
Integrating near the boundary $x_0$ for example one has
$$\int_{x_0 - 0} ^{x_0 + 0} \, dx' \partial_{x'} \left( \alpha_
+^2 - \alpha_-^2 + w_1^+ - w_1^- \right) = 0\eqno\eq$$
This implies that at the boundary $x = x_0 $ one has
$$\epsilon_+ (t,x_0 ) + \epsilon_- (t,x_0 ) = w_1^+ (t,x_0 )
- w_1^- (t,x_0 )\eqno\eq$$
and there is indeed an interaction between the two sets of fields.
More concretely the discrete modes of the topological field
are identified with the non--Dirichlet modes of the collective field
which are now allowed to exist.  In conclusion the collective field
$\alpha_{\pm}$ has Dirichlet modes
(vanishing at the boundary) whose dynamics is unchanged by the
additional fields.  In addition one now also has non--Dirichlet modes
which are related to the discrete degrees of freedom.
The question of the dynamics at the turning
points was raised recently [15] and we have now described how the
improved
collective field formulation deals with these issues.

The implication of
the above considerations on scattering processes and tunnelling
are left for future
investigations.

In general the presence of supplementary fields opens the
possibility for constructing general classical backgrounds of
string theory.  Currently we have considered the simplest ground
state solution of the coupled theory.  One can expect a more general
class of solutions given by the 3-dimensional KP equation.  Its
integrability structure will be of certain help in studying them.

\noindent{\bf Acknowledgements:}  J. A. wishes to thank Profs. O.
Babelon, M. V. Saveliev, P. Sorba and K. Takasaki for discussions on
2+1 Toda theory.  A. J. is grateful to Prof. M. Ninomiya and Prof. T.
Yoneya for their hospitality at the UJI Research Center and the
University of Tokyo (Komaba).

\endpage

\noindent\underbar{{\bf Appendix A}}

$W_N$ algebras were introduced as generalizations of the
infinite--dimensional conformal Virasoro algebra [22].  They are
naturally associated with KdV--type integrable systems.  Accordingly,
their canonical scheme of construction for the classical case is
the Feigin-Fuchs
construction [30] using a Toda field theory.  We describe here the
particular case of $WA_n$ (usually denoted $W_n $) algebras
associated
with $sl(n,{\bf C} )$ Toda field theories.
Introducing
the bosonic $N$-component field $\left\{ {\vec \phi}_i , i = 1 \dots
N\right\}$ with Poisson brackets
$$\left\{\phi_i\, ,\, \partial \phi_j\right\} = \delta (z - z^\prime\,
) K_{ij}\eqno(A.1)
$$
$K$ being the Cartan matrix of $sl (N,{\bf C})$, the $W_{N+1}$-algebra
is obtained (classically) by considering the differential operator:
$$
D = \prod_{c = 1}^{N+1} \> (\partial_z + h_\mu \cdot \partial
\phi)\eqno(A.2)
$$
$\{h_\mu\}$ being a basis of root vectors of $s\ell (N+1)$ completed by
$h_{N+1} = -\Sigma_1^N h_n$ [23].  Expanding $D$ as a $N+1$--order
differential
operator, its components denoted $u_i$ are quasi-primary fields of spin
$j$ (for the
term $\sim \partial^{N+1-j}$ ), and must then be deformed to
generate the $W_{N+1}$ classical algebra.  The Poisson structure [35]
for the
coefficients $u_n$ deduced from (A.1) and (A.2) is in fact the
second, or Gel'fand-Dikii Poisson structure, for the order $N+1$ KdV
Lax operator.  Choosing for $\{h_\mu\}$ a quasi-
orthonormal ``basis" $h_\mu \cdot h_\nu = \delta_{\mu \nu} - 1/N$ (in
order to respect $\Sigma_{\mu = 1}^{N+1}\> h_\mu = 0$), one has the
Poisson bracket algebra:
$$
\left\{h_\mu \cdot \partial \phi\, ,\>\> h_\nu \cdot \partial
\phi\right\} = (\delta_{\mu \nu} - 1/N) \> \partial_z \delta \, (z -
z^\prime\, )\eqno(A.3)$$
A realization of $h_\mu$ by particular $N$-dimensional vectors exists,
which triggers a remarkable recursion relation.  We choose [19]:
$$\eqalign{
h_1 &= \left ({1\over {\sqrt{2}}}\, ,\quad {1\over {\sqrt{6}}}\
, ,\,\dots
{1\over {\sqrt{i (i+1)}}}\, ,\, \dots {1\over {\sqrt{N (N+1)}}}
\right )\cr
\vdots &\cr
h_K &= \left (0 \dots 0;\>{-K+1\over {\sqrt{K(K-1)}}} \> ;\> {1\over
{\sqrt{K (K+1)}}} \>,\, \dots {1\over {\sqrt{N(N+1)}}}\right )\cr
\vdots &\cr
h_{N+1} &= \left ( 0 \qquad \dots \quad , 0 ,\, {-N\over {\sqrt{N
(N+1)}}}\right )}\eqno(A.4)$$
One sees on this expression that each first $K$ components of the
first $K+1$ vectors (denoted $h^{(k)}$) builds a similar Cartan dual
basis,
this time associated with the algebra $W_{K+1}$.  A recursive structure
then
follows, allowing to express systematically a $W_{N+1}$-algebra as a
combination of generators of a $W_N$ algebra, and polynomial
currents in a supplementary bosonic field $\phi_N (z, {\bar z}\, )$,
Poisson commuting with the previous bosonic fields $\phi_i (i= 1\dots
N-1)$.  We recall that:
$$
\left\{ (\partial\phi)_i (z),\>\> (\partial\phi)_j (z^\prime\,
)\right\} = \delta_{i,j} \>
\delta^\prime\, (z - z^\prime\, )\eqno(A.5)
$$
The currents $w_n$ are now obtained from improved
Feigin--Fuchs coefficients of (A.2).
For instance:
$$\eqalign{
w_2 &= u_2\cr
w_3 &= u_3 - {n-2\over 2} \>\partial u_2\cr
w_4 &= u_4 - c_1\>\partial u_3 - c_2\> \partial u_2 - c_3 \> u_2^2\cr
\dots &}\eqno(A.6)
$$
and reciprocally the Feigin-Fuchs coefficients are easily rewritten in
terms of the $w_n$ currents by a recursive inversion of (A.6):
$$\eqalign{
u_2 &= w_2\cr
u_3 &= w_3 + {n-2\over 2} \> \partial w_2\cr
u_4 &=w_4 - c_1 \>\partial w_3 - c_1 {n-2\over 2} \>\partial^2 w_2 -
c_2\>\partial^2 w_2 - c_3\>w_2^2}\eqno(A.7)
$$
We now write explicitly the $N+1$-order differential operator (A.2),
separating the coefficients containing the field $\phi_N$ and the
coefficients containing the $N-1$ first fields:
$$\eqalign{
D^{(N)} &= \left ( \partial - {N\over {\sqrt{N(N+1)}}} \> \partial
\phi_N\right ) \>\> \left ({\overleftarrow \prod}_{i=1}^N \quad
\partial + h_i^{(N-1)} \cdot \partial \phi + {\partial \phi_N\over
{\sqrt{N(N+1)}}}\right )\cr
&\equiv \left ( \partial - {N\over {\sqrt{N(N+1)}}} \partial
\phi_N\right ) \>\> {\hat D}^{(N-1)}}\eqno(A.8)
$$
The operator $\hat D^{(N-1)}$ is nothing but a gauge-transformed
version of the operator $D^{(N-1)}$ generating the $W_{N-1}$ algebra:
$$
{\hat D}^{(N-1)} = e^{- \phi_N /\sqrt{N(N+1)}} \cdot D^{(N-1)} \cdot
e^{\phi_N /\sqrt{N(N+1)}}\eqno(A.9)
$$
Defining:
$$
j_n \equiv e^{- \phi_N /\sqrt{N(N+1)}} \cdot \partial^n\cdot\> e^{\phi_N
/\sqrt{N(N+1)}}\eqno(A.10)$$
one has:
$$
\eqalign{
{\hat D}^{(N-1)} & =  \partial^N + \sum_{n=1}^N {\hat u}_n^{(N-
1)} \>\> \partial^{N-n}\cr
{\hat u}_n^{(N-1)} &= u_n^{(N-1)} + \sum_{p=1}^n  j_p (\phi_N) \>
u_{n-p}^{(N-1)} }\eqno(A.11)
$$
Now:
$$\eqalign{
D^{(N)} & = \left( \partial + {N\over {\sqrt{N(N+1)}}} \partial
\phi_N\right)\cdot \hat{D}^{(N-1)}\cdot\cr
& = \partial^{N+1} + \sum_{n=1}^{N+1} \left( {\hat u}_n^{(N-1)} +
\partial
{\hat u}_{n-1}^{(N-1)} + {N\over {\sqrt{N(N+1)}}} \partial \phi_N
\cdot \hat{u}_{n-1}^{(N-1)}\right)\partial^{N-n+1} }\eqno(A.12)
$$
where it is understood that $\hat{u}_{N+1}^{(N-1)} \equiv 0$ and $\hat
{u}_0^{(N-1)} \equiv 1$.  From (A.12), (A.10) and (A.7), it naturally
follows that:
$$\eqalign{
u_n^{(N)} = \,& u_n^{(N-1)} +\> \left[{\rm terms\>in}\>u_{p<n}^{(N-
1)}\right] + \,\,[{\rm coupling\,\, terms}\cr
& {\rm for} \,\, u_{p<n}^{(N-1)} \, {\rm and} \,\, j
(\phi_N)] + [ {\rm
pure\,\,terms\,\,in} \,\,\phi_N ] }\eqno(A.13)$$

In particular the pure $\phi_N$-terms are easily obtained as:
$$u_n^{(N)} [\phi_N ] = j_n (\phi_N ) + \partial j_{n-1} (\phi_N ) +
{N\over{\sqrt{N(N+1)}}} \, \partial \phi_N . j_{n-1} (\phi_N
)\eqno(A.14)$$

Finally, from (A.6), it follows that the currents generating the
classical $W_N$--algebra from the Feigin--Fuchs construction can be
expressed as a coupling between currents generating a $W_{N-1}$
algebra and currents $j_n (\phi_N)$ of a single bosonic field $\phi_N$.
Moreover, if the central charge of the Virasoro algebra $\{w_2\}$ is
made to vanish by a suitable choice of improved energy--momentum
tensor, the $W_{N-1}$ algebra can be consistently set to zero and one
ends up with a $1$--field realization of a centerless $W_N$ algebra.
The limit $N\rightarrow \infty$ of this realization can be identified
with the standard bosonization of $W_{1+\infty}$ mentioned in section
2.

Notice that this procedure generalizes the $W_3$ ``factorized"
construction described in [21].

\noindent\undertext{{\bf Appendix B}}

We recall here the relations between the large $N$ KdV Lax operator
and the $w_{\infty}$--generating currents, described in [9].
Starting from the Lax operator:

$${\cal L} = \partial^N + u_2 \, \partial^{N-2} + \cdots u_N\eqno(B.1)$$
where the coefficients $u_N \cdots u_2$ obey the second Gelfand--Dikii
Poisson algebra  [35] one introduces a long--wavelength generating
functional where $\partial_x$ is replaced by a c--number $\lambda^{-
1}$:

$${\cal L}  = \lambda^{-N} \left( 1 + u (\lambda )\right) \quad ;
\quad u(\lambda ) \equiv  \sum_{s=2}^N \, u_s \lambda^s \eqno(B.2)$$

Define then

$${\cal L}^{1/N} = \lambda^{-1} \left( 1 + u (\lambda) \right)^{1/N} =
\lambda^{-1} \left( 1 + {1\over N} \ln (1 + u (\lambda )) + \cdots
\right) \eqno(B.3)$$

In the notation of [9], $\ln (1 + u (\lambda )) \equiv w(\lambda )$ which
generates the $w_{\infty}$ densities as $w (\lambda ) =
\sum_{s=2}^{\infty} \lambda^s w_s$.  But simultaneously, the conserved
Hamiltonians of the long wavelength KdV hierarchy become:

$$\eqalign{ h^n & = \oint d\lambda \cdot \lambda^{-2} \, {\cal
L}^{n/N}\cr
& = \oint d\lambda \cdot \lambda^{-2} (\lambda^{-n} ) \left( 1 + {1\over
N} w (\lambda ) + 0 (1/N^2 )\right) ^n\cr
& = \oint d\lambda\cdot \lambda^{-n-2} \left( 1 + {nw(\lambda)\over N}
+ 0
(1/N^2) \right) \cr
& = {n\over N} \, w_{n+1} + 0 (1/N^2 ) }\eqno(B.4)$$

This proves the statement that the generating currents of the
$w_{\infty}$ algebra, in the large $N$ KdV framework, are the
hamiltonian densities of the KdV hierarchy in its long wavelength
limit.

\noindent\undertext{{\bf Appendix C}}

We shall here detail some recently understood features of the
construction in [26] of a $w_{\infty}$ current algebra.
We first recall that the continuum field variable
$\partial\partial_s\Phi^{{\rm Toda}}$ must be identified with the
potential $(\vec{h}_n \cdot \partial\vec{\phi})$ in the differential
operator defining the Feigin--Fuchs construction (see Appendix A,
formula (A.2)).  Indeed this
potential arises in the Toda-Lax representation [27] as the
$n$--th term on
the diagonal of the Lax matrix.  Notice that Bakas [9] uses a
different set of field variables, namely contracting $\phi_i$ by the
inverse of the Killing metric $k^{ab}$, leading to the Toda equation:
$$\partial\bar{\partial} \, \phi_i = \sum_{\{\alpha\}} \, \alpha_i
\cdot \exp
(\alpha\cdot \phi ) \Rightarrow \partial\bar{\partial} (k\phi ) = \exp
(k\phi )\eqno(C.1)$$

It follows that this field variable must be identified with the
inverse--derivative of our continuum field variable, hence $\phi_{{\rm
(Bakas)}} = \Phi^{{\rm Toda}}$, in a perfectly consistent way.

Now one can explicitely check the continuum limit of the first non--
trivial current density, namely $w_3$.  We recall that its continuum
value is [26]:
$$w_3 = \int ds \, \bigl\{ - {(\partial\partial_s \phi )^3\over 3} + 2
(\partial\partial_s \phi ) \partial^2 \phi + s (\partial^3 \phi +
\partial^2 \phi \, \partial\partial_s^2 \phi )\bigl\} \eqno(C.2)$$

The discrete value of the third component in the Feigin--Fuchs
construction follows from (A.2) and properties of the root vectors
$h_{\mu}$ [23]:

$$\eqalign{ u_3 & = \sum_{\mu} \, {(h_{\mu} \partial\phi )^3\over 3}
- \sum_{\mu} (h_{\mu} \partial \phi ) (N - \mu) (h_{\mu} \partial^2 \phi ) -
\sum_{N\geq \nu} \, (h_{\mu} \partial \phi) (h_{\nu} \partial^2
\phi)\cr
& + \sum_{\mu} \, (h_{\mu} \partial \phi ) (h_{\mu}
\partial^2 \phi ) + \sum_{\mu} \, {-2N+1\over 2} \, \mu h_{\mu}
\partial^3 \phi + \sum {\mu^2\over 2} \, h_{\mu} \partial^3
\phi }\eqno(C.3)$$

The exact $w_3$ current is in fact $u_3 - {n-2\over 2} \partial u_2
$.  When the proper normalization of $w$--currents, adapted to
the $w_{\infty}$ limit, is adopted, the terms $\sim N\partial w_2$
drop out since they are reduced by a factor $N^{-1/2}$ with respect to the
term $w_3$.  Dropping accordingly the corresponding terms in (C.3)
(namely the linear term in $N$ which is exactly $N \, \partial u_2$)
and moreover eliminating residual subleading terms like $\sum \mu
h_{\mu}\partial^3 \phi$ and $\sum h_{\mu} \partial \phi\cdot h_{\mu}
\partial^2
\phi$ one ends up with:
$$\eqalign{u_3 = & \sum_{\mu} \, {(h_{\mu}
\partial\phi )^3\over 3} + \sum_{\mu} \, (h_{\mu} \partial\phi )
(h_{\mu} \partial^2 \phi )\cdot \mu - \sum_{n\geq \nu} \, (h_{\mu}
\partial \phi) (h_{\nu} \partial^2 \phi)\cr
& + \sum {\mu^2\over 2} h_{\mu} \, \partial^3 \phi + {\rm
subleading \,\, orders} } \eqno(C.4)$$

The continuum limit of $u_3$, replacing $\sum_{\mu}$
by $\int ds$ and $(h_{\mu} \partial\phi )$ by $(\partial\partial_s
\phi )$, gives precisely (C.2).  This check is a supplementary
confirmation of the validity of the expression (C.2), previously
established [26] by computing non--trivial Poisson brackets; moreover
it relates directly the Feigin--Fuchs (Gelfand--Dikii) construction
to the Lax representation of $2+1$ Toda theory.

Note that the vanishing of the central charge in the Poisson
bracket $\{ w_2 (s),\, w_2(s' ) \}$ for the $2+1$ continuous theory
follows from the definition of the ``continuous" limit of the sum
$\sum_{\mu}$ as an
Adler--type residue integral $\oint ds \times \cdots $ [28].  Indeed the
central terms in the discrete $w_{\infty}$ algebra arose from
sums over the roots of $A_{n-1}$, of the form $\sum_{\mu} 1, \mu,
\mu^2$; their continuous limits $\oint ds, sds, s^2ds$ are automatically
zero under the Adler trace convention, hence the continuous $w_{\infty}$
algebra has a vanishing central charge.

\endpage

\noindent{\bf References}

\pointbegin
I. Klebanov; ``String Theory and Quantum Gravity '91 ", Trieste
Spring School 1991, edited by J. Harvey et al.; p. 30-102, {\it World
Scientific} (1992).
\point
A Jevicki, B. Sakita; {\it Nucl. Phys.} {\bf B165} (1980), 511; S. R.
Das, A. Jevicki; {\it Mod. Phys. Lett.} {\bf 5} (1990), 1639.
\point
J. Polchinski; {\it Nucl. Phys.} {\bf B 346} (1990), 253 and {\bf
B362} (1991), 25.
\point
K. Demeterfi, A. Jevicki, J. Rodrigues; {\it Nucl. Phys.} {\bf B362}
(1991), 165 and {\bf B365} (1991) 489.
\point
E. Br\'ezin, Cl. Itzykson, G. Parisi, J. B. Zuber; {\it Comm. Math.
Phys.} {\bf 39} (1978), 35.
\point
D. Gross, I. Klebanov; {\it Nucl. Phys.} {\bf B352} (1991), 671;
A. M. Sengupta, S. Wadia, {\it Intern. Journ. Mod. Phys.}
{\bf A6} (1991), 1961;
G. Moore; {\it Nucl. Phys.} {\bf B368} (1992), 557.
\point
J. Avan, A. Jevicki; {\it Phys. Lett.} {\bf B266} (1991), 35 and {\bf
B272} (1990), 17; J. Avan, A. Jevicki; to appear in {\it Comm. Math.
Phys.}(1992).
\point
D. Minic, J. Polchinski, Z. Yang; {\it Nucl. Phys.} {\bf B369} (1992),
324; M. Awada, S. J. Sin; UFIFT (Florida)-HEP 90-33 and 91-03; G.
Moore,
N. Seiberg; {\it Int. Journ. Mod. Phys.} {\bf A7} (1992) 2601;
S. R. Das, A. Dhar, G. Mandal, S. Wadia; {\it Mod. Phys.
Lett.} {\bf A7} (1992), 71.
\point
I. Bakas; {\it Phys. Lett.} {\bf B228} (1989), 57; {\it Comm. Math.
Phys.} {\bf 134} (1989), 487; C. Pope, L. Romans, X.Shen; {\it Phys.
Lett.} {\bf B236} (1990), 173.
\point
D. Gross, I. Klebanov and M. Newman; {\it Nucl. Phys.} {\bf B350}
(1991), 621.
\point
A. M. Polyakov, {\it Mod. Phys. Lett.} {\bf A6} (1991), 635;
Preprint PUPT (Princeton)-
1289 (Lectures given at 1991 Jerusalum Winter School).
\point
E. Witten; {\it Nucl. Phys.} {\bf B373} (1992), 187;
I. Klebanov, A. Polyakov; {\it Mod. Phys. Lett.} {\bf A6} (1991),
3273; I. Klebanov; {\it Mod. Phys. Lett.} {\bf A7} (1992), 723; N.
Sakai, Y. Tanii; {\it Prog. Theor. Phys.} {\bf 86} (1991),
547.
Y. Matsumura, N. Sakai, Y. Tanii; TIT (Tokyo) --HEEP 127,186 (1992).
\point
J. Avan and A. Jevicki; {\it Mod. Phys. Lett.} {\bf A7} (1992), 357.
\point
S. Iso, D. Karabali and B. Sakita; CCNY-HEP 92-1 (1992), To Appear in
{\it Nucl. Phys.} {\bf B}; CCNY-HEP-92-6.
\point
A. Dhar, G. Mandal and S. Wadia, TIFR-TH-91/61 (1992).
\point
M. V. Saveliev; {\it Comm. Math. Phys.} {\bf 121} (1989), 283; M.V.
Saveliev, A. V. Vershik; {\it Comm. Math. Phys.} {\bf 126} (1989),
367; R. M. Kashaev et al. in ``Ideas and Methods in Mathematical
Analysis", ed. by S. Albeverio et al., Cambridge Univ. Press (1992).
\point
S. Tomonaga; {\it Prog. Theor. Phys.} (Kyoto) {\bf 5}, 544 (1950).
\point
M. Fukuma, H. Kawai, R. Nakayama; {\it Comm. Math. Phys.} {\bf 143}
(1992), 371.
\point
H. Lu, C. N. Pope; {\it Phys. Lett.} {\bf B286} (1992), 63.
\point
X. Shen, X. J. Wang; Preprint CPT-TAMU (Texas A\&M) 59-91 (1991).
\point
L. J. Romans; {\it Nucl. Phys.} {\bf B352} (1991), 829.
\point
V. A. Fateev, A. B. Zamolodchikov; {\it Nucl. Phys.} {\bf B280} (1987),
644; V. A. Fateev, S. L. Lykyanov; {\it Intern. Journ. Mod. Phys.}
{\bf A3} (1988), 507.
\point
A. Bilal, J. L. Gervais, {\it Nucl. Phys.} {\bf B314} (1989), 646 and
{\bf B318} (1989), 579;
O. Babelon; {\it Nucl. Phys. B} (suppl. {\bf 18A}) (1990), 1; A.
Bilal, `` String Theory and Quantum Gravity '91", see ref. 1; M.
Bershadski, H. Ooguri; {\it Comm. Math. Phys.} {\bf 126} (1989), 49.
\point
K. Takasaki, T. Takebe; {\it Lett. Math. Phys.} {\bf 23} (1992), 205;
K. Takasaki; KUCP (Kyoto) 0049-92.
\point
Q-Han Park; {\it Phys. Lett.} {\bf B236} (1992), 429.
\point
J. Avan; {\it Phys. Lett.} {\bf A168} (1992), 363.
\point
I. Leznov, M. V. Saveliev; {\it Lett. Math. Phys. } {\bf 3} (1979),
489; {\it Phys. Lett.} {\bf B79} (1978), 294; O. Bogoyavlensky; Izv.
Akad. Nauk. SSSR (Math.) {\bf 52} (1988), 712.
\point
M. Adler; {\it Invent. Math.} {\bf 50} (1979), 219; G. Segal, G.
Wilson; Publications IHES {\bf 61} (1985), 5; E. Date et al.;
``Transformation Groups in Soliton Equations", in ``Non--linear
Integrable Systems", ed. by M. Jimbo and T. Miwa, {\it World
Scientific} (1983).
\point
P. Sorba, M. V. Saveliev, private communication. See also 14.
\point B. L. Feigin, D. B. Fuchs; {\it Funct. Anal. Appl.} {\bf 16}
(1982), 114; {\bf 17} (1983), 241.
\point
A. Bilal; {\it Phys. Lett.} {\bf B227} (1991), 406.
\point
A. M. Kirillov, ``Theory of Group Representation", ed. by Mir (French
version)(1974).
\point
J. Matsukidarai, J. Satsuma, W. Strampp; {\it Journ. Math. Phys.} {\bf
31} (1990), 1426.
\point
V. G. Drinfel'd, V. I. Sokolov; {\it Journ. Sov. Math.} {\bf 30}
(1984), 1975; A. Jevicki, T. Yoneya; {\it Mod. Phys. Lett.} {\bf A5}
(1990), 1615.
\point
I. M. Gel'fand, L. A. Dikii; Preprint 136, IPM.A.N.SSSR (1978),
unpublished.
\point
E. Witten; {\it Comm. Math. Phys.} {\bf 121} (1989), 351; X. G. Wen,
{\it Int. Journ. Mod. Phys.} {\bf B4} (1990), 239.
\point
I. I. Kogan; UBC-TP-92-23 (1992).
\endpage
\centerline{{\bf Fig. Caption 1}}
\vskip .50in
\item{} Space Domains for the Collective (straight line) and
Topological (wavy line) Fields in a $V = -x^2$ potential.
\bye